\documentclass[lettersize,journal,review]{IEEEtran}
\usepackage{amsmath,amsfonts}
\usepackage{algorithmic}
\usepackage{algorithm}
\usepackage{array}
\usepackage{stfloats}
\usepackage{url}
\usepackage{verbatim}
\usepackage{graphicx}
\graphicspath{{Figures/}}
\usepackage{cite}
\usepackage{graphicx}
\usepackage{array}
\usepackage{booktabs}
\usepackage{pifont}
\usepackage{cleveref}
\usepackage{subfig}
\usepackage[font=small]{caption}

\usepackage[numbers,sort&compress]{natbib}
\begin{document}

\title{Spatial-Temporal Transformer based Video \\ Compression Framework}

\author{Yanbo Gao, Wenjia Huang, Shuai Li\textsuperscript{*},~\IEEEmembership{Senior Member,~IEEE}, Hui Yuan,~\IEEEmembership{Senior Member,~IEEE}, \\Mao Ye, Siwei Ma,~\IEEEmembership{Senior Member,~IEEE}
        % <-this % stops a space
\thanks{This work was supported in part by the National Key R\&D Program of China under Grant 2018YFE0203900; in part by the National Natural Science Foundation of China under Grant 62271290 and Grant 62001092; in part by the Natural Science Foundation of Shandong Province under Grant ZR2022ZD38 and in part by SDU QILU Young Scholars Program. }%\textsuperscript{*}Corresponding Author: Shuai Li}
\thanks{Y. Gao and W. Huang are with School of Software, Shandong University, and S. Li and H. Yuan are with School of Control Science and Engineering, Shandong University, Jinan, China. E-mail: \{ybgao,shuaili\}@sdu.edu.cn. \par M. Ye is with School of Computer Science and Engineering, University of Electronic Science and Technology of China, Chengdu, China.  E-mail: cvlab.uestc@gmail.com.\par S. Ma is with National Engineering Research Center of Visual Technology, School of Computer Science, Peking University, Beijing, China. E-mail: swma@pku.edu.cn.}
% <-this % stops a space
\thanks{Manuscript received Sep. 21, 2023.}}

\markboth{IEEE Transactions on Circuits and Systems for Video Technology}%
{Shell \MakeLowercase{\textit{et al.}}: A Sample Article Using IEEEtran.cls for IEEE Journals}


\maketitle

\begin{abstract}
Learned video compression (LVC) has witnessed remarkable advancements in recent years. Similar as the traditional video coding, LVC inherits motion estimation/compensation, residual coding and other modules, all of which are implemented with neural networks (NNs). However, within the framework of NNs and its training mechanism using gradient backpropagation, most existing works often struggle to consistently generate stable motion information, which is in the form of geometric features, from the input color features. Moreover, the modules such as the inter-prediction and residual coding are independent from each other, making it inefficient to fully reduce the spatial-temporal redundancy.  To address the above problems, in this paper, we propose a novel Spatial-Temporal Transformer based Video Compression (STT-VC) framework. It contains a Relaxed Deformable Transformer (RDT) with Uformer based offsets estimation for motion estimation and compensation, a Multi-Granularity Prediction (MGP) module based on multi-reference frames for prediction refinement, and a Spatial Feature Distribution prior based Transformer (SFD-T) for efficient temporal-spatial joint residual compression. Specifically, RDT is developed to stably estimate the motion information between frames by thoroughly investigating the relationship between the similarity based geometric motion feature extraction and self-attention. MGP is designed to fuse the multi-reference frame information by effectively exploring the coarse-grained prediction feature generated with the coded motion information.  SFD-T is to compress the residual information by jointly exploring the spatial feature distributions in both residual and temporal prediction to further reduce the spatial-temporal redundancy. Experimental results demonstrate that our method achieves the best result with 13.5\% BD-Rate saving over VTM and 68.7\% BD-Rate saving over the baseline without the proposed modules. Ablation study validates the effectiveness of each proposed module.
\end{abstract}


\section{Introduction}\label{sec1}
\IEEEPARstart{V}{ideo} data has experienced an exponential growth with the proliferation of video-sharing platforms and increasing high-resolution videos, resulting in the urgent need of more efficient video compression. Video compression is to reduce the spatial and temporal redundancy with intra/inter prediction, transform coding, quantization, and entropy coding, in order to achieve high compression ratios while maintaining perceptual quality.

Traditional block-based video coding approaches have been widely studied with predefined intra/inter predictions such as the angular intra prediction and motion estimation/compensation, predefined transform coding such as the DCT, and predefined entropy coding such as the CABAC \cite{Bross2021OverviewOT}. Such block-based video coding architecture with predefined modules has enjoyed a great success in the past decades and widely adopted in the industry. With the rapid development of deep learning, there has been a growing interest in exploring new video compression methods based on deep neural networks.  These methods aim to leverage the powerful representation learning capabilities of deep models to learn adaptive transforms instead of using predefined ones, in order to achieve higher compression efficiency. Various approaches have been proposed  \cite{Hu2021FVCAN,Li2023NeuralVC,Liu2020NeuralVC,Hu2022CoarseToFineDV, Lin2020MLVCMF,Wang2023ButterflyMR,Yi2022TaskDrivenVC, Li2018IndependentlyRN,Lin2023DMVCDM,Chen2021MOVICodecDV, Mentzer2022VCTAV,Zou2021LearnedVC, Zhang2023VideoCA,Arnab2021ViViTAV,Liu2021VideoST,Cao2021VideoST, Sheng2021TemporalCM,Zou2020EndtoEndLF,Li2021DeepCV} including deep learning enhanced video coding by replacing some modules in the traditional video coding method and learned video coding approach with a whole neural network to compress a video. This paper focuses on the learned video coding, especially learned inter-frame video coding. Existing learned inter-frame video coding approach generally takes a similar process as the traditional video coding, including inter-frame prediction based on motion estimation/compensation, residual compression and entropy coding. While many learned methods \cite{Li2022HybridSE,Li2023NeuralVC} have been developed and achieve state-of-the-art performance as traditional video coding method, there still exist two key problems in the exploration of the inter-frame information.

Firstly, motion information as geometric information, used to align the reference frame to the current frame in order to perform the inter-frame prediction, is difficult to be stably transformed from the color space, i.e., the image frame and its corresponding feature. Current alignment methods mainly use optical flow and offsets as motion, and both are dense geometric transformation representations. Learning from color representations based on gradient backpropagation to generate such geometric feature is usually not stable as illustrated in \cite{Hu2021FVCAN}. Moreover, existing methods mostly only employ the immediate previous frame for prediction without fully exploring the multiple reference frames. 

Secondly, after the inter-frame prediction, the residual is compressed independently from the inter-frame prediction information, neglecting the useful spatial information in the prediction. To be specific, other than subtracting the corresponding point-to-point prediction information as a temporal prediction, the spatial relationship embedded in the prediction can also assist the spatial compression of the residual.

To address the above problems, a spatial-temporal Transformer based inter-frame video coding framework is proposed in this paper. First, a Multi-Granularity Prediction generation with proposed Relaxed Deformable Transformer is developed, where the multi-reference frame information is fully explored. Then a Spatial Feature Distribution prior based Transformer (SFD-T) is proposed to utilize the spatial feature distribution prior embedded in the temporal prediction feature to reduce the remaining spatial redundancy in the residual. The contributions of this paper can be summarized as follows.

\begin{itemize}
\item{We propose a Relaxed Deformable Transformer (RDT) based motion estimation and compensation module, where the RDT transforms the color feature with the spatial position embedding to generate the geometric motion alignment information and produces a coarse prediction feature. The mechanism of using RDT for producing motion between two frames is thoroughly investigated, and the deformable transformer is relaxed to the deformable convolution with their relationship carefully examined.}
\item{We propose a Multi-Granularity Prediction (MGP) based multi-reference prediction refinement module. The multi-reference frame information is explored in the manner of video denoising with the coarse prediction feature as anchor, since it is obtained with coded motion information and thus contains most information of the current frame.}
\item{We propose a Spatial Feature Distribution prior based Transformer (SFD-T) module to compress the residuals by further exploring the spatial feature distribution information embedded in the temporal prediction.}
\end{itemize}

The above modules are all constructed in the form of transformer and comprises a complete Transformer based video coding framework. Extensive experiments demonstrate that our method achieves state-of-the-art results, and ablation studies have also been conducted to validate the proposed modules.

The rest of this paper is organized as follows. Section \ref{sec2} presents the related works in learned image and video compression. Section \ref{sec3} describes the proposed method in details. Experimental results with ablation studies are provided in Section \ref{sec4} and conclusions are drawn in Section \ref{sec5}.

\section{Related Work}\label{sec2}
In this section, a brief review of the related work in the field of learned video compression  \cite{Hu2021FVCAN,Li2023NeuralVC,Liu2020NeuralVC,Hu2022CoarseToFineDV, Lin2020MLVCMF,Wang2023ButterflyMR,Yi2022TaskDrivenVC, Li2018IndependentlyRN,Lin2023DMVCDM,Chen2021MOVICodecDV, Mentzer2022VCTAV,Zou2021LearnedVC, Zhang2023VideoCA,Arnab2021ViViTAV,Liu2021VideoST,Cao2021VideoST, Sheng2021TemporalCM,Zou2020EndtoEndLF,Li2021DeepCV} is presented. Considering that image compression usually serves to code the first frame of a video and the spatial motion and residual used in the video coding, some related learned image compression methods are reviewed first.

\subsection{Learned Image Compression}
Existing learned image compression methods are mostly derived from Variational Autoencoder (VAE) \cite{Ball2016EndtoendOI}, which uses an Encoder-Decoder architecture with the quantization process mimicked as the variational process. Most of the encoders and decoders adopt the convolutional neural networks (CNNs) \cite{LeCun1989BackpropagationAT} with down-sampling and up-sampling in the encoder and decoder, respectively. The input is first transformed into the latent representation with the encoder, and quantized by rounding which is replaced by adding a uniform-distributed noise for gradient backpropagation at training  \cite{Ball2018VariationalIC}. The quantized residual is then encoded into the bitstream with the entropy coding and transmitted to the decoder  \cite{Hu2021FVCAN,Ball2016EndtoendOI,Lu2021TransformerbasedIC}. To enhance the performance of entropy coding, hyperprior and context based entropy coding methods  \cite{Lu2021TransformerbasedIC,Zou2022TheDI,Koyuncu2022ContextformerAT} were developed and widely used in the following learned image and video compression. 

With the rapid development of Transformers \cite{Vaswani2017AttentionIA,Dosovitskiy2020AnII}, Transformer based image compression has also been studied \cite{Lu2021TransformerbasedIC,Zou2022TheDI,Koyuncu2022ContextformerAT}. A CNN-Transformer hybrid framework of image compression was proposed in  \cite{Lu2021TransformerbasedIC},  where the Swin Transformer blocks are inserted between convolutional layers without changing the overall architecture. In \cite{Zou2022TheDI}, a symmetric Transformer framework is used by replacing all CNN layers in the encoder-decoder architecture with Transformer layers. In addition to exploring Transformer for the encoder and decoder, there are also works such as ContextFormer \cite{Koyuncu2022ContextformerAT} and Entroformer \cite{Qian2022EntroformerAT} investigating using Transformer for entropy model.

These researches focus on reducing the spatial redundancy within an image, and can also be used in video coding to compress the motion vectors (if exist) and residuals. However, these models are developed without considering the temporal information and cannot directly using them for video coding. 

\subsection{Learned Video Compression}
Compared with the image compression, learned video compression focuses more on the inter-frame predictive coding. Most of the existing methods adopt a similar procedure as the conventional hybrid video coding, and consist of motion estimation and compensation  \cite{Liu2020NeuralVC,Hu2022CoarseToFineDV,Lin2020MLVCMF,Yi2022TaskDrivenVC}, residual coding and filtering \cite{Zou2020EndtoEndLF, Wang2023ButterflyMR}. Deep video compression (DVC)  \cite{Lu2018DVCAE} first proposed a fully learned video compression, using optical flow and warping to represent motion and perform motion compensation to generate prediction frame. Then residual is obtained by subtracting the current frame with the prediction frame, and further coded with an autoencoder (similarly as the image compression). Many methods have been developed with a similar procedure. In  \cite{Hu2021FVCAN}, FVC proposed to perform the inter-prediction in the feature domain using deformable convolution to reduce the artifacts around edges brought by the optical flow. Deep Contextual Video Compression (DCVC) \cite{Li2021DeepCV},  and its variants, including DCVC-TCM \cite{Sheng2021TemporalCM}, DCVC-HEM \cite{Li2022HybridSE} and DCVC-DC \cite{Li2023NeuralVC}, were proposed to use context models for spatial compression. It directly processes the concatenated prediction and the current frame without explicitly obtaining the residual, using the property of conditional entropy no larger than the entropy of the residual.       

In addition to the single-reference based prediction methods described above, there also exist methods using multi-scale or multi-reference frame strategy  \cite{Hu2022CoarseToFineDV,Lin2020MLVCMF,Liu2020NeuralVC} to help with prediction generation. In \cite{Wang2023ButterflyMR}, the multi-reference frames are first gradually warped to the current frame with the coded motion vectors at each time step, and then the warped multi-reference frames are fused together to generate the final prediction. 3D convolution is used in \cite{Yi2022TaskDrivenVC} to fuse the initial prediction feature and the multi-reference frames without temporal warping. Other than explicitly fusing multi-reference frames at each frame, implicit temporal information aggregation with neural architectures such as LSTM  \cite{Hochreiter1997LongSM} have also been investigated for video coding \cite{Li2018IndependentlyRN}. In \cite{Lin2023DMVCDM}, Conv-LSTM is used together with the U-net architecture to directly learn the space-time difference between video frames. LSTM have also been explored to construct the context models for the hyperprior based entropy coding  \cite{Wang2023ButterflyMR,Chen2021MOVICodecDV}.

\begin{figure*}[t] %H为当前位置，!htb为忽略美学标准，htbp为浮动图形
\centering %图片居中
\includegraphics[width=1\textwidth]{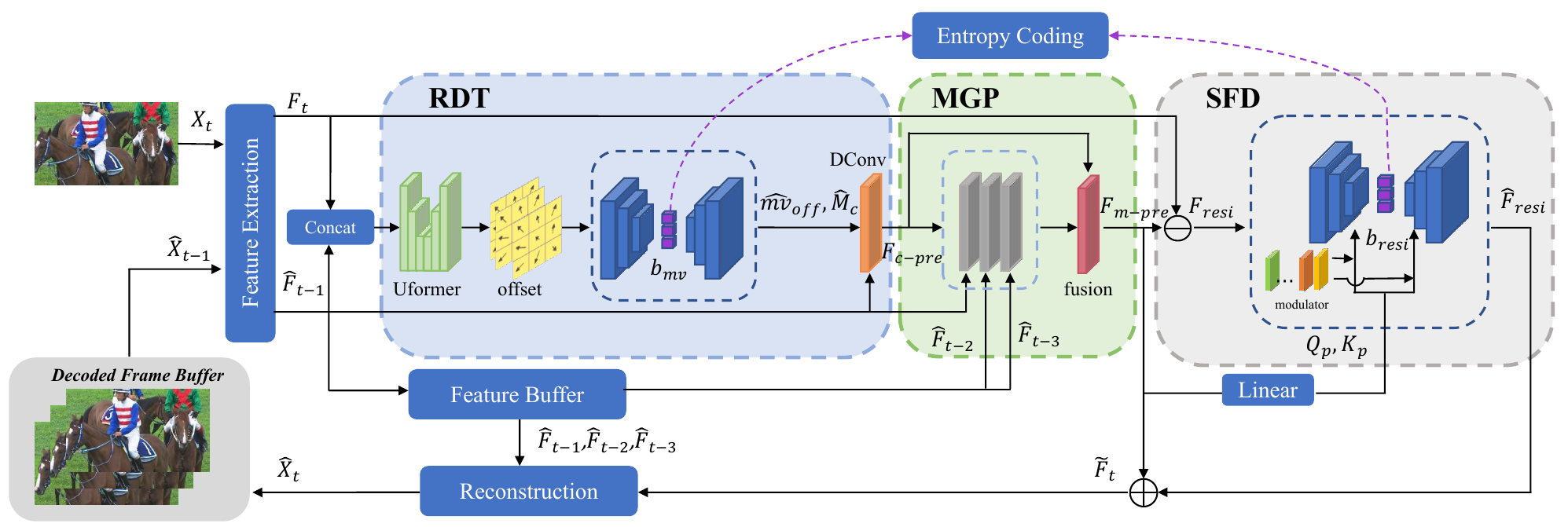} %插入图片，[]中设置图片大小，{}中是图片文件名
\caption{Framework of the proposed STT-VC.} %最终文档中希望显示的图片标题
\label{fig_1} %用于文内引用的标签
\end{figure*}

The above methods all employ the CNNs based encoding architecture. On the other hand, while there are some Transformer based spatial-temporal modelling methods, there only exist few Transformer based video coding methods. VCT \cite{Mentzer2022VCTAV} directly used Transformer to model temporal dependencies and predict the distribution of  current frame without using explicit motion representation. Autoregressive prediction within each frame is also used similarly as the autoregressive context model \cite{Minnen2018JointAA}.  Some methods also adopt the self-attention module in part of the video coding framework such as in encoder and decoder \cite{Zou2021LearnedVC} and quality enhancement  \cite{Zhang2023VideoCA,Zou2020EndtoEndLF}. 

In the area of Transformer based temporal modelling, several representative modelling methods are briefly described in the following. ViVit \cite{Arnab2021ViViTAV} proposed a Transformer based video processing method by factorizing the model into spatial and temporal processing. For the temporal processing, patches of the same spatial position are grouped with a temporal position embedding to synthesize three-dimensional tokens and thus can be processed in the same way as the spatial Transformer. Video SwinTransformer \cite{Liu2021VideoST} extended the Swin-T \cite{Liu2021SwinTH} to three dimensions for video processing, where a 3D shifted window over a video is used for the multi-head self-attention calculation. CRAFT \cite{Sui2022CRAFTCF} was developed for optical flow estimation based on  cross-attention  between two frames. VSRT \cite{Cao2021VideoST} proposed a spatially integrated convolutional self-attention layer to enhance the extraction of spatial locality information and use an optical flow based temporal alignment to enhance temporal information modelling. Deformable Transformer was first proposed for spatial object detection \cite{Zhu2020DeformableDD}, where the offsets and the weights are generate by a query feature and then the sampled features are fused together similarly as the deformable convolution. Then Deformable Attention Transformer (DAT) \cite{Xia2022VisionTW}  was proposed, where, after the offsets are generated, the sampled features are fused together using the self-attention model. However, such Transformer based temporal modelling methods do not concern the special needs of video coding such as the balance between rate and distortion, and cannot be directly used for video coding.

\section{Proposed Method}\label{sec3}

\subsection{Overview}
This paper proposes a learned inter-frame video coding method to generate a high-quality prediction and a compactly coded residual, with a Transformer based framework. It mainly consists of three components, including Relaxed Deformable Transformer (RDT) for motion estimation and compensation, Multi-Granularity Prediction (MGP) for prediction refinement, and Spatial Feature Distribution prior based Transformer (SFD-T) for residual compression. The framework of the proposed method is illustrated in Fig. \ref{fig_1}.

The current frame ${X_t}$ and reference frame ${\hat{X}_{t-1}}$ are first transformed to features ${F_t}$ and ${\hat{F}}_{t-1}$, respectively, and the following prediction and residual compression are performed in the feature domain. 1) RDT first uses Uformer to conduct the motion estimation (shown in the light blue box of Fig. \ref{fig_1}), which is then coded with a motion vector codec and finally produces the motion information ${\widehat{mv}}_{off}$ and attention scores/confidence mask ${\hat{M}}_c$. Then the deformable Transformer based value feature fusion process is relaxed to the deformable convolution to generate a coarse-grained prediction feature ${F_{c-pre}}$. 2) MGP is further used to refine the prediction feature to explore the multi-reference frame information (shown in the light green box of Fig. \ref{fig_1}). MGP applies RDT to align the multi-reference frame features to the coarse-grained prediction feature ${F_{c-pre}}$, and fuses them through a spatial-channel attention to generate enhanced prediction feature ${F_{m-pre}}$.  The residuals ${F_{resi}}$ are generated by subtracting the enhanced prediction feature ${F_{m-pre}}$ from the current frame feature ${F_t}$. 3) SFD-T compresses the residual ${F_{resi}}$ (shown in the light gray box of Fig. \ref{fig_1}), by exploring the enhanced prediction feature ${F_{m-pre}}$ in the attention calculation process as a spatial feature distribution prior. Finally, the decoded residual feature ${{\hat{F}}_{resi}}$ is added with the enhanced prediction feature ${ F_{m-pre}}$ to reconstruct input feature ${{\hat{F}}_t}$, which is then further transformed back to pixel domain as the reconstructed frame ${{\hat{X}}_t}$. The three proposed modules RDT, MGP and SFD-T are presented in detail in the following subsections.

\begin{figure}[t] %H为当前位置，!htb为忽略美学标准，htbp为浮动图形
\centering %图片居中
\includegraphics[width=0.48\textwidth]{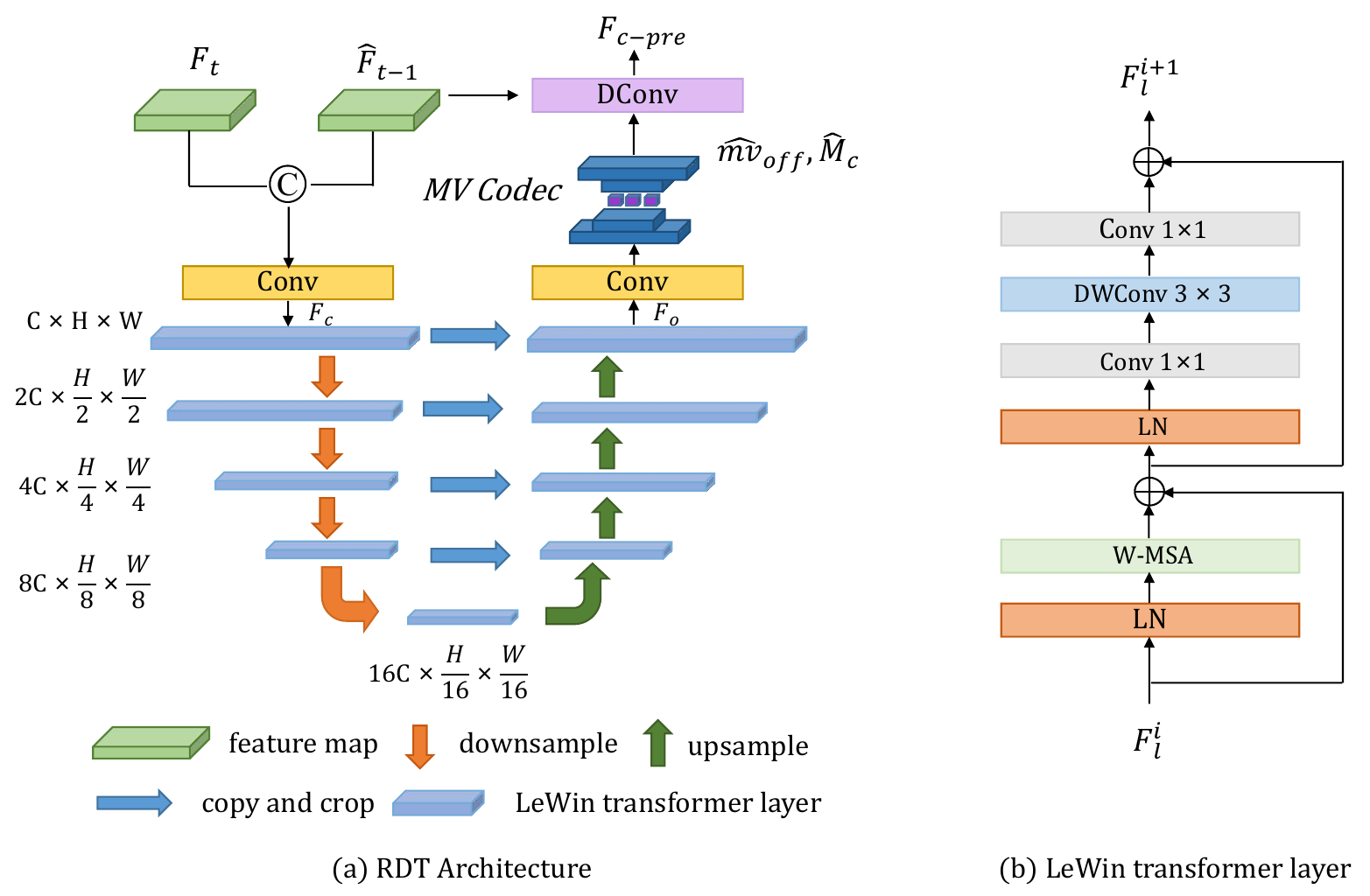} %插入图片，[]中设置图片大小，{}中是图片文件名
\caption{(a) Structure of the proposed RDT based motion estimation and compensation. (b) Details of the Lewin transformer layer with W-MSA \cite{Liu2021SwinTH} used in RDT. %最终文档中希望显示的图片标题
\label{fig_2}} %用于文内引用的标签
\end{figure}

\subsection{Relaxed Deformable Transformer (RDT) for Motion Estimation and Compensation}
To perform inter-frame prediction, the motion information between the current frame and the reference frame needs to be estimated first, known as motion estimation. Then the reference frame is aligned to the current frame with the estimated motion information to generate prediction, known as motion compensation. Currently, warping with optical flow and using deformable convolution \cite{Dai2017DeformableCN} with offsets are the two main approaches for motion estimation and compensation. It is known that utilizing optical flow for feature alignment during motion compensation often results in artifacts around the edges of objects. Therefore, in this paper, offsets with the deformable convolution (DConv) is used as the base to represent the motion and perform motion compensation between the reference frame and the current frame. 

The existing offsets and DConv based motion estimation and compensation methods usually estimate the offsets by gradually convolving the features of the reference and current frames using a CNN. However, the offsets, in the form of the geometric information, is difficult to be directly obtained from the color features via gradient backpropagation. On the other hand, motion estimation and compensation estimates the motion changes between the reference frame and the current frame based on their similarity, and then aligns the reference frame with the motion to ensure its similarity to the current frame. Essentially this process is conducted based on the similarity between the reference frame and the current frame. Therefore, to overcome the problem of stably obtaining similarity based geometric motion information from color features, a relaxed deformable transformer (RDT) is developed. A Uformer is first used to estimate the offsets and attention/confidence of each offsets based on the similarity, and a deformable convolution is then used as a relaxed deformable Transformer to fuse the value features according to the corresponding offsets and attention/confidence. 

First, the current frame and the reconstructed reference frame are transformed into the feature domain through a feature extraction module. Taking the processing of the current frame $X_t$ as an example, it can be represented by
\begin{align}
F_{conv}=ReLU({Conv}_{5\times5}(X_t))\nonumber\\%\label{deqn_ex1}\\
F_t=ResBlocks(F_{conv})+F_{conv}\label{deqn_ex2}
\end{align}
where $F_t$ represents the final feature of $X_t$. $ResBlocks$ represent three Resblocks, each of which consists two convolution modules and a skip connection.

To calculate the similarity between the features of the reference frame and current frame, instead of using cross-attention \cite{Sui2022CRAFTCF}, we find that simply concatenating the features and processing them together with the self-attention can better cover the similarity calculation among different positions of the two features and within each feature. Specifically, the two frame features $F_t$ and ${\hat{F}}_{t-1}$ are concatenated together, and processed with a ${1\times1}$ convolution to fuse the information and reduce the channels as 
\begin{align}
F_c={Conv}_{1\times1}\langle F_t,{\hat{F}_{t-1}} \rangle \label{deqn_ex3}
\end{align}
where $\left \langle .\right \rangle$ is used to represent the concatenation operation for simplicity. Then the spatial position embedding is incorporated into the features and processed by the self-attention model as
\begin{align}
F_s=W_{MSA}(LN(F_c+F_{pos}))\nonumber\\%\label{deqn_ex4}\\
W_{MSA}=softmax({QK}^T/\sqrt d+B)V\label{deqn_ex5}
\end{align}
where ${W_{MSA}}$ represents the calculation of window-based multi-head self-attention  \cite{Liu2021SwinTH}, and ${LN}$ represents the normalization layer. ${F_{pos}}$ represents the spatial position embedding, which is obtained by embedding the absolute position within a block. ${Q,K,V }$ represent the query, key and value features in the calculation of multi-head self-attention, obtained by conducting linear projection to features at each layer (${Q,K,V=\ Linear(LN(F_c+F_{pos}))}$), and $B$ represents relative position bias as in  \cite{Liu2021SwinTH}.

Since the concatenated feature ${F_c}$ contains information from both ${F_t}$ and ${F_{t-1}}$, the calculation of attention scores ${(Q\cdot K)}$ before normalization between features of two positions ${\left(x,y\right)}$ and\ ${(x+\Delta x,y+\Delta y)}$ can be formulated as the following Eq. (\ref{deqn_ex6}) by substituting  ${F_c }$ with Eq. (\ref{deqn_ex3}) and ignoring the position embedding first.

\vspace{-0.3cm}
%\begin{align}
%F\left(x,y\right)\cdot & F(x+\Delta x,y+\Delta y)=\notag\\
%& \quad  [f_a(F_t(x,y))+f_b(F_{t-1}(x,y))]\notag\\
%&\cdot[f_a(F_t(x+\Delta x,y+\Delta y))\notag\\
%&+f_b(F_{t-1}(x+\Delta x,y+\Delta y))]\label{deqn_ex6}
%\end{align}
\begin{small}
\begin{align}
F(x,&y) \cdot F(x+\Delta x,y+\Delta y)= [f_a(F_t(x,y))+f_b(F_{t-1}(x,y))]\notag\\
&\cdot[f_a(F_t(x+\Delta x,y+\Delta y))+f_b(F_{t-1}(x+\Delta x,y+\Delta y))]\label{deqn_ex6}
\end{align}
\end{small}         
where ${f_a}$ and ${f_b}$ represent the information mixing of ${F_t}$ and ${F_{t-1}}$ in the ${1\times1}$ convolution in Eq. (\ref{deqn_ex3}). Note that here we do not concern the detailed function of  ${f_a}$ and ${f_b}$ and simply use them to represent the information fusion process. Eq. (\ref{deqn_ex6}) can be further turned into
\begin{align}
F\left(x,y\right)\cdot &\ F(x+\Delta x,y+\Delta y)=\notag\\
&\quad \
  f_a(F_t(x,y))\cdot f_a(F_t(x+\Delta x,y+\Delta y))\notag\\
&+f_a(F_t(x,y))\cdot f_b(F_{t-1}(x+\Delta x,y+\Delta y))\notag\\
&+f_b(F_{t-1}(x,y))\cdot f_a(F_t(x+\Delta x,y+\Delta y))\notag\\
&+f_b(F_{t-1}(x,y))\cdot f_b(F_{t-1}(x+\Delta x,y+\Delta y))
\label{deqn_ex7}
\end{align}

It can be seen that the self-attention score calculation on the fused features covers the estimation of the similarity not only within each feature but also between features, which is important for motion estimation.

On the other hand, position embedding introduces the spatial geometric information into the features. In the above self-attention score calculation, with the position embedding, it can learn to also consider the geometric distance among different positions in addition to the feature similarity. More importantly, it helps directly transfer the color features into geometric information. After the attention score calculation, the value features ($V$) are fused together based on the attention score. Since $V$ is obtained with the position embedding, the final feature can be represented as
\begin{align}
F_s=\sum_{i\in B}\alpha_iV(F_c+F_{pos})\label{deqn_ex8}
\end{align}
where ${ i\in B}$ represents the block (${\{8\times8\}}$ in the experiments) of the self-attention calculation. It can be seen that by incorporating the similarity information contained in the attention score and the geometric information contained in the value feature, the output feature directly contains the desired geometric position information based on the similarity among features. This agrees with the motion estimation based on the similarity as discussed in the beginning of this subsection. Thus, the motion offsets between two features can be stably obtained by the above self-attention.

To obtain the motion information which distributes in a large range from small motion of a fractional pixel to large motion of a few pixels, the Uformer architecture \cite{Wang2021UformerAG} is used. It directly calculates the self-attention based on pixel-wise features within a block instead of patchifying them, thus able to obtain detailed motion. On the other hand, to achieve the large motion based on global information, U-structure is used by down-sampling and up-sampling the features, and also concatenating the encoder feature to the output with a skip connection. In each layer of the Uformer, in addition to the self-attention calculation as in Eq. (\ref{deqn_ex5}) the input feature is also added to the output feature ${F_a=F_s+F_c}$. Then a few convolutional layers are further used to increase the local processing of the features, known as locally-enhanced window (LeWin) Transformer layers \cite{Wang2021UformerAG}.

\vspace{-0.3cm}
\begin{small}
\begin{align}
&F_l={{Conv}_{1\times1}(DWConv}_{3\times3}({Conv}_{1\times1}(LN(F_a))))+F_a \label{deqn_ex9}
\end{align}
\end{small}
where $DWConv$ represents depth-wise convolution. By using LeWin Transformer layers, it effectively models the relationship between pixels in the reference frame and the current frame within each window. The other layers in the U-architecture use the same form and produce the final output feature $F_o$ as shown in Fig. \ref{fig_2}. Finally, a $1\times1$ convolution can be used in the end to transform the output feature into motion offsets between the current frame and the reference frame. Since in video coding the motion information needs to be compressed into the bitstream and transferred to the decoder, the final feature of Uformer, instead of the motion offsets, is compressed with an encoder-decoder codec and then transformed into the motion offsets with the decoder feature. This motion encoding process can be expressed as
\begin{gather}
F_{mv-o}=Enc(F_o) \nonumber\\%\label{deqn_ex10}\\
b_{mv}=Entropy({f_Q(F}_{mv-o})) \nonumber\\%\label{deqn_ex11}\\
{\widehat{mv}}_{off},{\hat{M}}_c={Conv}_{1\times1}(Dec({\hat{F}}_{mv-o})) \label{deqn_ex12}
\end{gather}
\begin{figure}[t] %H为当前位置，!htb为忽略美学标准，htbp为浮动图形
\centering %图片居中
\includegraphics[width=0.5\textwidth]{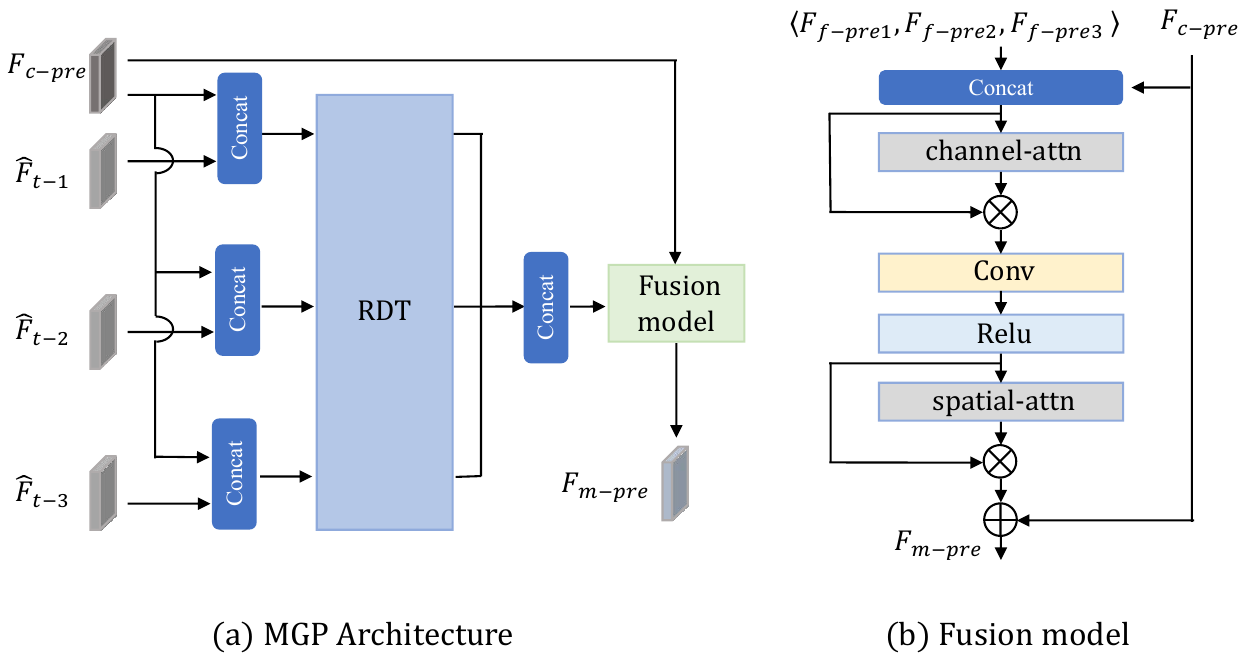} %插入图片，[]中设置图片大小，{}中是图片文件名
\caption{(a) Structure of the proposed MGP. (b) Details of the fusion model used in MGP.} %最终文档中希望显示的图片标题
\label{fig_3} %用于文内引用的标签
\end{figure}
where $Enc$ and $Dec$ represent the encoder and decoder in the motion codec, respectively. Any motion codec can be used and here for simplicity, the motion codec in FVC \cite{Hu2021FVCAN} is used. $f_Q$ represents the quantization process generating the quantized feature ${\hat{F}}_{mv-o}$, and $Entropy$ refers to the entropy coding model. Finally, the decoded features are processed with a $1\times1$ convolution to generate the reconstructed motion offsets ${{\widehat{mv}}_{off}}$.  Unlike the conventional deformable transformer that uses the deformed features indicated by the motion offsets to enhance the current feature with the self-attention operation \cite{Xia2022VisionTW}, here the deformed features are fused together to generate a prediction of the current feature. Therefore, an attention mask ${{\hat{M}}_c}$ is directly obtained together with the motion offsets as a relaxed version of self-attention.
\begin{align}
F_{c-pre}=\sum_{i\in\{3\ast3\}}&{{{\hat{M}}_c(i)Conv}_{1\times1}({\hat{F}}_{t-1}((x_i,y_i)+{\widehat{mv}}_{off}(i)))}\notag\\
&{\xrightarrow{relax}}{DConv}_{3\times3}({\hat{F}}_{t-1},{\widehat{mv}}_{off},{\hat{M}}_c)		
\label{deqn_ex13}
\end{align}
where ${{\hat{F}}_{t-1}(\left(x_i,y_i\right)+{\widehat{mv}}_{off}(i))}$ represents the deformed reference frame features (bilinearly interpolated), and ${Conv}_{1\times1}$ is $1\times1$ convolution to generate the value feature. Summarizing the value features with the attention $ {\hat{M}}_c\left(i\right)$ over the block produces the final prediction feature. This process is equivalent to a deformable convolution (DConv) with shared filter weights over the $3\times3$ locations, where ${\widehat{mv}}_{off}$ and ${\hat{M}}_c$ represent the offsets and confidence mask, respectively, and the multi-head number in the transformer is the group number in the DConv. For generality, we further relax the shared embedding weights to non-shared weights as convolution, and thus turning it to a deformable convolution. This process, deformable convolution with offsets and confidence mask obtained with a Uformer, is thus coined to relaxed deformable transformer (RDT), unifying the motion estimation and compensation in the realm of Transformer.

\subsection{Multi-granularity prediction (MGP) feature generation based on Multi-frame Enhancement}
After motion estimation and compensation, a coarse- grained prediction feature is generated. However, using only the immediately previous frame cannot provide accurate prediction of the current frame, especially for moving areas and occluded background. Predicting such information requires the long-range temporal information which can be partly provided by the multi-reference frames. With the pixel-wise motion representation for prediction generation in the existing LVC framework, it is difficult to directly generate a prediction by mixing the motion from multi-reference frames, or explore the multi-reference frames without significantly increasing the rate on the motion representation. On the other hand, the motion representation is obtained under the rate-distortion optimization (RDO), where, instead of obtaining a motion representation of a high rate to provide the best-quality prediction feature, a motion representation with the smallest RD cost is used to provide a decent-quality prediction with a relatively small rate. Therefore, the motion representation is suboptimal only considering the quality of prediction feature. 

To solve the above problems, a multi-granularity prediction generation method is proposed to fully explore the multi-reference frames. It is developed based on the observation that the coarse-grained prediction feature, generated with the motion offsets encoded to the bitstream, contains much information of the current frame and can be considered as a noisy version of the current frame. Unlike the previous multi-reference prediction methods focusing on predicting the motion vectors of multi-reference frames or directly fusing the past multiple prediction features based on the past noisy motion information, we propose to take the coarse-grained prediction feature as an approximate of the current frame feature. Accordingly, the multi-reference frames are explored in a manner of video denoising where the motion information used in the process can be obtained at both encoder and decoder without coding into bitstream.

\begin{figure}[t] %H为当前位置，!htb为忽略美学标准，htbp为浮动图形
\centering %图片居中
\includegraphics[width=0.5\textwidth]{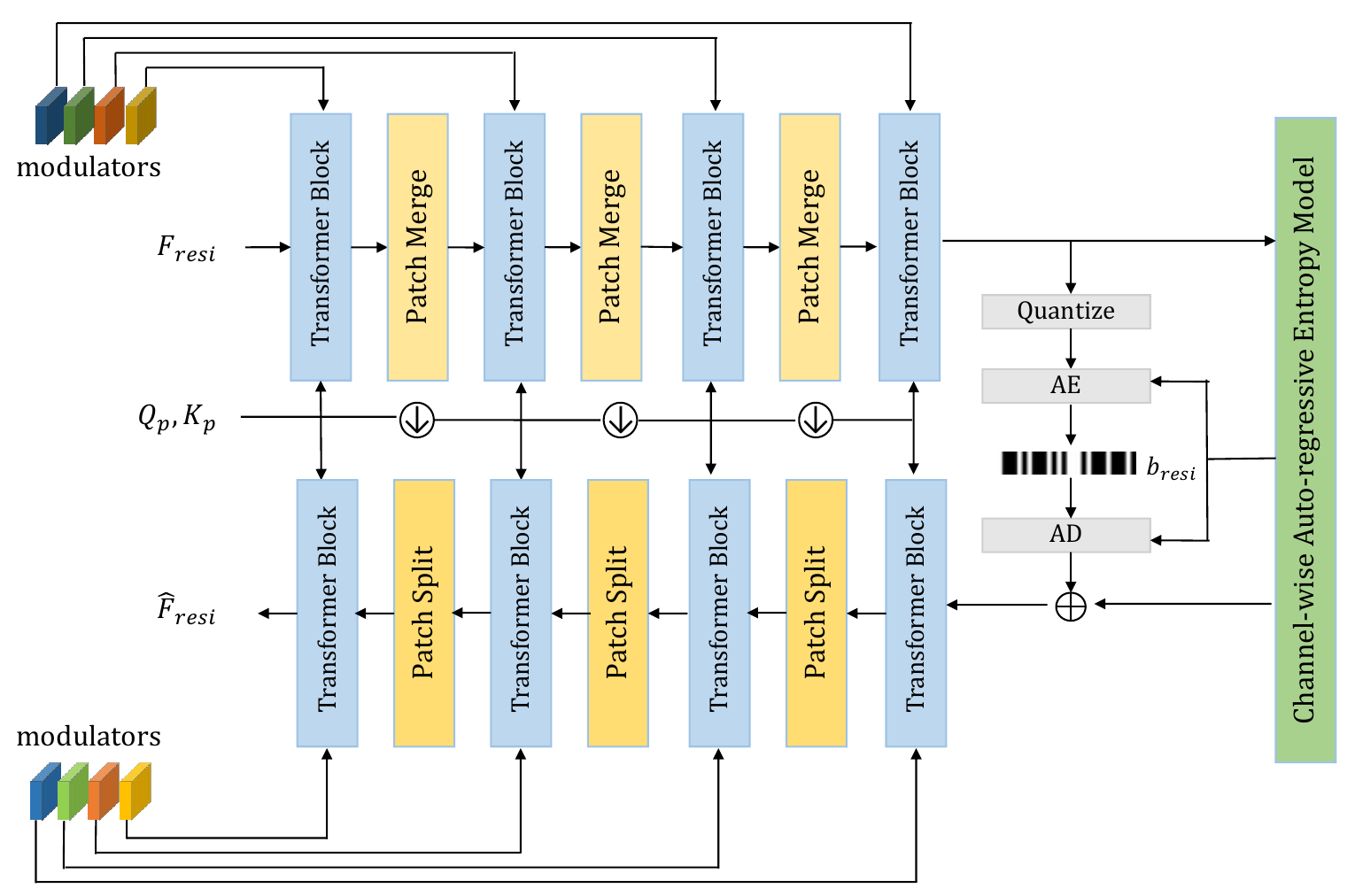} %插入图片，[]中设置图片大小，{}中是图片文件名
\caption{Structure of the proposed SFD-Transformer based residual compression.} %最终文档中希望显示的图片标题
\label{fig_4} %用于文内引用的标签
\end{figure}

The features of the multiple reference frames, including the immediately previous frame feature ${\hat{F}}_{t-1}$, are first temporally aligned to the prediction feature $F_{c-pre}$ and then fused together with the prediction feature to improve its quality. Taking ${\hat{F}}_{t-1}$ as an example, the temporal alignment process is similar to the above coarse-grained prediction feature generation:
\begin{align}
F_{f-pre1}=RDT (\langle F_{c-pre},\hat{F}_{t-1}\rangle )\label{deqn_ex14}
\end{align}
where $RDT$ represents the above Relaxed Deformable Transformer. Note that here the motion representation is not coded with quantization and entropy coding since the coarse-grained prediction feature $F_{c-pre}$ and the reference frame feature ${\hat{F}}_{t-1}$ are available at both encoder and decoder. Therefore, the motion representation encoded under RDO in the coarse-grained prediction feature generation only needs to describe the coarse motion to save bits, while the detailed motion representation and prediction feature can be further generated with this bit-free temporal alignment and enhancement process. The same process is applied to the other reference features to explore the long-range temporal information. In the experiments, three reference frames are used with features ${\hat{F}}_{t-1}$, ${\hat{F}}_{t-2}$ and ${\hat{F}}_{t-3}$, and corresponding enhanced features using RDT are denoted by $F_{f-pre1}$, $F_{f-pre2}$ and $F_{f-pre3}$, respectively. Such enhanced features can be considered as fine-grained prediction feature, which is generated without the rate cost of the motion representation. 

The final prediction feature is obtained by fusing the coarse-grained prediction feature and the multi-reference fine-grained prediction feature. It is known that frames that are closer in the temporal dimension tend to have higher similarity. As a result, each reference feature contributes differently to the prediction feature, indicating that the importance of  $F_{f-pre(i)}$ varies. Moreover, in the spatial domain, the distribution of image details differs between flat regions and sharp edges, and also between moving regions and background, leading to variations in the spatial fusion of the different features. To address this, a spatial-channel attention similar as CBAM \cite{Woo2018CBAMCB} is used to fuse the features with different channel and spatial weights. This fusion process is illustrated in Fig. \ref{fig_3} and can be described as
\begin{align}
F_{enh\_cat}&= \langle F_{c-pre},F_{f-pre1},F_{f-pre2},F_{f-pre3} \rangle \nonumber\\%\label{deqn_ex15}\\
F_{attn\_ch}&=C\_attn(F_{enh\_cat})\cdot F_{enh\_cat}\nonumber\\%\label{deqn_ex16}\\
F_{enh\_conv}&=ReLU({Conv}_{1\times1}(F_{attn\_ch}))\nonumber\\%\label{deqn_ex17}\\
F_{attn\_sp}&=S\_attn(F_{enh\_conv})\cdot F_{enh\_conv}\nonumber\\%\label{deqn_ex18}\\
F_{m-pre}&=F_{c-pre}+F_{attn\_sp}\label{deqn_ex19}
\end{align}
where $C\_attn$ and $S\_attn$ represents the channel and spatial attention in CBAM \cite{Woo2018CBAMCB}, respectively. The $1\times1$ convolution reduces the channel number of $F_{attn\_ch}$ to the same as $F_{c-pre}$. The final prediction feature $F_{m-pre}$ is obtained by adding the coarse-grained prediction feature and the fused feature since the quality of the coarse-grained prediction feature is rather stable with the estimated and encoded motion representation. Finally, the residual feature is obtained by subtracting the current frame feature with the enhanced prediction feature $F_{resi}{=F}_t-F_{m-pre}$.

\subsection{Spatial Feature Distribution (SFD) Prior based Residual Compression}
After the residuals generated with the inter-prediction, residual compression is further performed to remove the spatial redundancy within the residual. Existing LVC methods preform residual compression completely in the spatial dimension, neglecting the inherent spatial feature distribution information that may be contained in the prediction frame. Taking an image with spatial repetitive patterns as an example, the spatial feature distribution in the residual, including the feature similarity at different locations, still resembles the distribution of the prediction features. In other words, the current inter-prediction only removes the pixel-to-pixel temporal redundancy with the subtraction operation or locally processed context operation, while the redundancy of the spatial feature distribution between temporal frames is ignored and can be further reduced. Therefore, a Spatial Feature Distribution prior based Transformer (SFD-Transformer) is developed for residual compression, where the spatial prior presented in the prediction feature is used to guide the self-attention model in compressing the spatial residual features.

\begin{table*}[t]
\renewcommand{\arraystretch}{1.2} % 可以调节, 1.2指高度是默认的1.2倍
\caption{Result comparison in terms of BD-Rate (\%) measured with PSNR. The anchor is VTM.}\label{tab:table1}
\centering
\begin{tabular}{lrrrrrrr}
%\begin{tabular}{m{2.5cm}<{} m{1.5cm}<{\centering} m{1.5cm}<{\centering} m{1.5cm}<{\centering} m{1.5cm}<{\centering} m{1.5cm}<{\centering} m{1.5cm}<{\centering} m{1.5cm}<{\centering} }
\toprule
\makebox[2.5cm][l]{Method} & \makebox[1.5cm][r]{UVG} & \makebox[1.5cm][r]{MCL-JCV} & \makebox[1.5cm][r]{HEVC B} & \makebox[1.5cm][r]{HEVC C} & \makebox[1.5cm][r]{HEVC D} & \makebox[1.5cm][r]{HEVC E} & \makebox[1.5cm][r]{Average} \\ \hline
VTM-17.0 \cite{VTM}  & 0.0 & 0.0 & 0.0 & 0.0 & 0.0 & 0.0 & 0.0 \\%\hline
HM-16.25 \cite{HM}  & 36.4 & 41.5 & 38.8 & 36.0 & 33.7 & 44.0 & 38.4 \\%\hline
ECM-5.0 \cite{ECM} & -10.0 & -12.2 & \textbf{ -11.5} & -13.4 & \textbf{-13.5} & -10.9 & -11.92 \\%\hline
CANF-VC \cite{Ho2022CANFVCCA} & 73.0 & 70.8 & 64.4 & 76.2 & 63.1 & 118.0 & 77.6 \\%\hline
DCVC \cite{Li2021DeepCV} & 166.1 & 121.6 & 123.2 & 143.2 & 98.0 & 266.1 & 153.0 \\%\hline
DCVC-TCM \cite{Sheng2021TemporalCM}  & 44.1 & 51.0 & 40.2 & 66.3 & 37.0 & 82.7 & 53.6 \\%\hline
DCVC-HEM \cite{Li2022HybridSE} & 1.1 & 8.6 & 5.1 & 22.2 & 2.4 & 20.5 & 10.0 \\%\hline
FVC \cite{Hu2021FVCAN} & 155.0 & 171.6 & 176.5 & 182.41 & 164.7 & 273.9 & 171.4\\\hline
PROPOSED & \textbf{-20.4} & \textbf{-16.4} & -10.1 & \textbf{-16.0} & -2.2 & \textbf{-15.8} & \textbf{-13.5} \\
\bottomrule
\end{tabular}
\end{table*}

The framework of the proposed SFD-Transformer is shown in Fig. \ref{fig_4}. An encoder-decoder Transformer architecture is used for the residual encoding and decoding. Firstly, at encoder, the residual is divided into patches, and embedded through a linear layer. Then the SFD prior based self-attention is calculated. Specifically, in the calculation of self-attention scores, the relationship among the prediction features is also considered via a feature distance as 

\vspace{-0.4cm}
{\begin{small}
\begin{align}
{\rm Attn}_{st-r}=softmax(Q_RK_R+{\rm pos}_b+Q_pK_p+mod)V_R  \label{deqn_ex20}
\end{align}
\end{small}}
where $Q_R$ and $K_R$ represent the query and key features of the current frame, respectively, while $Q_P$ and $K_P$ represent those of the corresponding prediction features, obtained with the same linear embedding functions. The similarity, i.e., $Q_P$ $K_P$, is served as the spatial feature distribution prior. ${\rm pos}_b $ and $mod$ represent the two-dimensional relative position coding and the learnable adjustment modulator, respectively. This updated self-attention calculation is used in all the Transformer layers at both encoder and decoder. $Q_p$ and $K_p $ at decoder are the same to the encoder, providing the same spatial feature distribution information for decoding the residual.

With reference to the existing Transformer based image compression model configuration \cite{Zou2022TheDI}, the number of self-attention blocks of each layer at the encoder is set to [x2,x2,x6,x2]. The resolution is reduced by down-sampling after each stage. It can be expressed as 

\vspace{-0.4cm}
\begin{small}
\begin{align}
{F_{resi(i)}}=f_{PM}({\rm Attn}_{st-r(i-1)}(F_{resi\left(i-1\right)},F_{m-pre(i-1)})) \label{deqn_ex21}
\end{align}
\end{small}
where ${\rm Attn}_{st-r\left(i-1\right)}()$ refers to the SFD prior based self-attention calculation operation of the ($i-1$)th layer. $f_{PM}$ represents the patch merging to down-sample the feature in the spatial resolution \cite{Liu2021SwinTH}, by reconstructing the feature between patches to be half in size and double in channel. Accordingly, $Q_p$ and $K_p $ are also down-sampled layer by layer with linear projection to match the size of corresponding residual feature maps, so that the SFD prior fits in the self-attention calculation of each layer. Take $Q_{p(i)}$ as an example,
\begin{align}
Q_{p(i)}=LN(Linear(Reshape(Q_{p(i-1)},\frac{1}{2}))) \label{deqn_ex22}
\end{align}
where $Reshape(Q_{p(i-1)},\frac{1}{2})$ as in  \cite{Wang2021PyramidVT}  represents reshaping the input $ Q_{p(i-1)}$  to the size $ \frac{HW}{4}\times\left(4C\right)$, and $Linear()$ is a linear projection that reduces channel number. $K_{p(i)}$ is processed in the same way.

The encoded feature $F_{resi(i)}$ after the encoder then undergoes entropy encoding to generate the bitstream.
\begin{equation}
\label{deqn_ex23}
b_{resi}=Entropy({f_Q(F}_{resi-o}))
\end{equation}

The bitstream is transmitted to the decoder and entropy decoding is performed to generate the initial residual feature. Then SFD prior based Transformer layers are used to further decode the initial residual features to reconstruct the residual features of the original resolution ${\hat{F}}_{resi}$. The prediction feature $F_{m-pre}$ is then added back to reconstruct the input feature.
\begin{equation}
\label{deqn_ex24}
{\widetilde{F}}_t=F_{m-pre}+{\hat{F}}_{resi}
\end{equation}

Similarly as FVC \cite{Hu2021FVCAN}, non-local attention mechanism is further used to enhance the reconstructed feature with the multi-reference reconstructed features. Finally, the enhanced reconstructed feature is transformed back to the pixel domain to generate the reconstructed frame ${\hat{X}}$  with a few ResBlocks in the same way as in FVC. 

The loss function of the proposed method is
\begin{equation}
\label{deqn_ex25}
L\ =R+\lambda D=R_{mv}+\ R_{resi}+\lambda d(X_t,{\hat{X}}_t)
\end{equation}
where $R_{mv}$ and $R_{resi}$ represent the bits introduced by compressing the offsets map ${\rm mv}_{off}$ and residual feature $F_{resi}$, respectively. $d(X_t,{\hat{X}}_t)$ is the distortion between original frame $X_t$ and reconstructed frame ${\hat{X}}_t$.  $\lambda$ is the corresponding Lagrange multiplier. For the first 15 epochs, the distortion $d(X_{m-pre},X_t)$ between enhanced prediction frame $X_{m-pre}$ (that is generated from the enhanced prediction feature $F_{m-pre}$ with a simple convolutional layer) and original frame $X_t$ is also used to accelerate the training of the RDT and MGP prediction module. After that, Eq.  (\ref{deqn_ex25}) is used to continue training.

\begin{table*}[t]
\renewcommand{\arraystretch}{1.2} % 可以调节, 1.2指高度是默认的1.2倍
\caption{Result comparison in terms of BD-Rate (\%) measured with MS-SSIM. The anchor is VTM.}\label{tab:table2}
\centering
\begin{tabular}{lrrrrrrr}
%\begin{tabular}{m{2.5cm}<{} m{1.5cm}<{\centering} m{1.5cm}<{\centering} m{1.5cm}<{\centering} m{1.5cm}<{\centering} m{1.5cm}<{\centering} m{1.5cm}<{\centering} m{1.5cm}<{\centering} }
\toprule
\makebox[2.5cm][l]{Method} & \makebox[1.5cm][r]{UVG} & \makebox[1.5cm][r]{MCL-JCV} & \makebox[1.5cm][r]{HEVC B} & \makebox[1.5cm][r]{HEVC C} & \makebox[1.5cm][r]{HEVC D} & \makebox[1.5cm][r]{HEVC E} & \makebox[1.5cm][r]{Average} \\ \hline
VTM-17.0 \cite{VTM} & 0.0 & 0.0 & 0.0 & 0.0 & 0.0 & 0.0 & 0.0  \\%\hline
HM-16.25 \cite{HM} & 31.1& 38.8& 36.6& 35.2& 33.0 & 41.1 & 36.0  \\%\hline
ECM-5.0 \cite{ECM} & -9.1 & -11.1 & -10.2 & -11.7 & -11.0 & -9.9 & -10.5 \\%\hline
CANF-VC \cite{Ho2022CANFVCCA}& 46.5 & 26.0 & 43.5 & 30.9 & 17.9 & 173.0 & 56.3 \\%\hline
DCVC \cite{Li2021DeepCV}  &64.9 & 27.5 & 54.4 & 39.7 & 15.2 & 210.4 & 68.7 \\%\hline
DCVC-TCM \cite{Sheng2021TemporalCM} & 1.0 & –10.8 & –11.7 & –15.2 & –29.0 & 16.7 & 8.85 \\%\hline
DCVC-HEM \cite{Li2022HybridSE}& –25.2 & –36.3 & –38.0 & –38.3 & \textbf{–48.1} & –25.8 & -35.3   \\%\hline
FVC \cite{Hu2021FVCAN}  & 144.8 & 151.9 & 150.8 & 119.9 & 116.2 & 244.7 & 154.7 \\\hline
PROPOSED  &\textbf{-36.0} & \textbf{-46.5} & -\textbf{43.1} & \textbf{-47.5} & -34.4 & -\textbf{36.8} & \textbf{-40.7} \\
\bottomrule
\end{tabular}
\end{table*}

\begin{figure*}[t] %H为当前位置，!htb为忽略美学标准，htbp为浮动图形
\centering %图片居中
\includegraphics[width=1\textwidth]{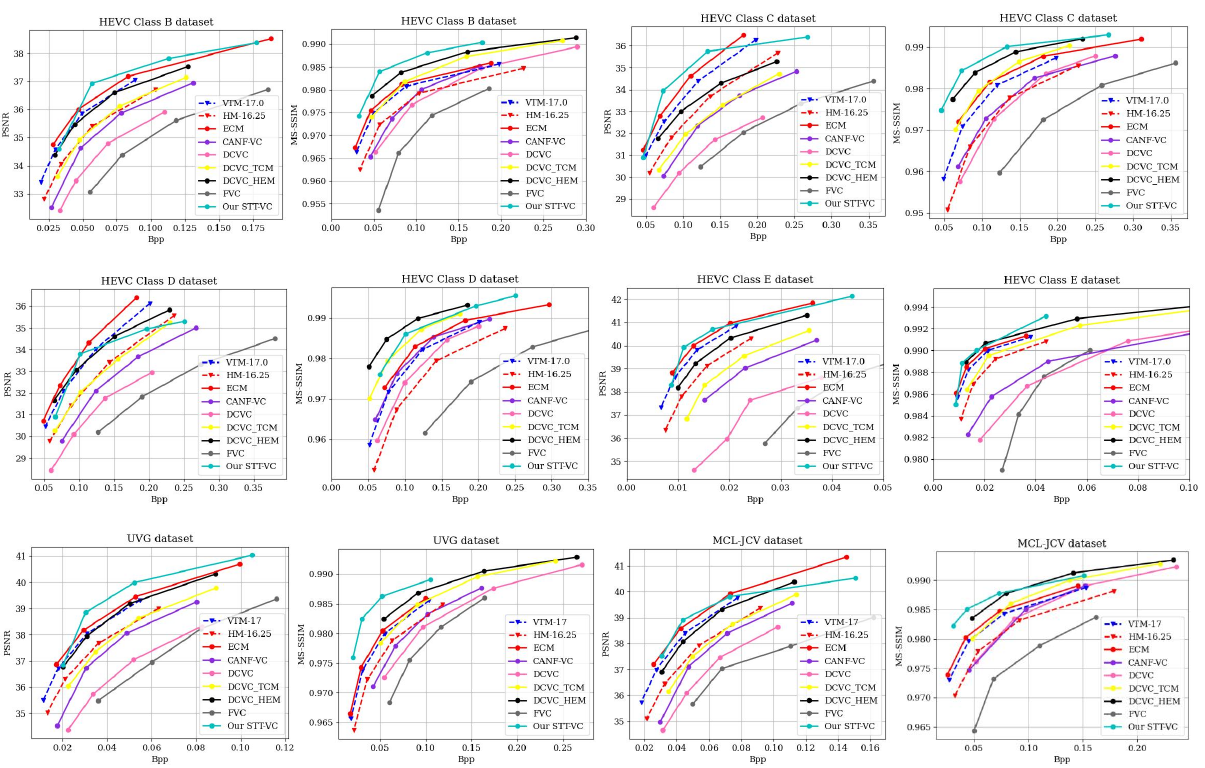} %插入图片，[]中设置图片大小，{}中是图片文件名
\caption{Rate-distortion curve comparisons under PSNR and MS-SSIM, respectively.} %最终文档中希望显示的图片标题
\label{fig_5} %用于文内引用的标签
\end{figure*}

\section{Experiments}\label{sec4}
\subsection{Experimental Settings}

\subsubsection{Dataset}
Vimeo-90K dataset \cite{Xue2017VideoEW}  is used as training dataset similar to the existing LVC methods \cite{Hu2021FVCAN,Sheng2021TemporalCM,Li2022HybridSE}. It contains 89,800 video sequences and each sequence includes 7 frames. The sequences are randomly cropped to the resolution of 256 × 256 as input for training. The first frame in GOP, namely I frame, is compressed by Cheng2020 in CompressAI \cite{Bgaint2020CompressAIAP}. Video sequences from three datasets are used for test to evaluate the performance of our model, including HEVC B-E class sequences \cite{Sullivan2012OverviewOT}, UVG \cite{Mercat2020UVGD5}, and MCL-JCV \cite{Wang2016MCLJCVAJ}.

\subsubsection{Training and testing Details}
The PyTorch platform is used for implementation. Adam \cite{Kingma2014AdamAM} optimizer is used with the batch size set to 8. The hyperparameter $\lambda$ is set to four different values corresponding to four models ($\lambda$ = 256, 512, 1024, 2048) at different rates. To construct the multi-reference frame structure in the proposed MGP, when the reference frame buffer has less than three frame features, the feature of the latest frame is duplicated until there are enough reference frames. Peak Signal-to-Noise Ratio (PSNR) and Multi-Scale-Structural Similarity Index (MS-SSIM) are used to evaluate the distortion between the reconstructed video and the ground-truth video. BD-Rate savings over both PSNR and MS-SSIM are adopted for evaluation and all evaluations are performed in the RGB space where the YUV videos are converted to RGB with FFMPEG. We evaluate 96 frames for each video in the test sets, with an intra period set to 32 frames. The other settings including the low-delay configurations are the same as in  \cite{Li2023NeuralVC}.

\begin{table*}[t]
\renewcommand{\arraystretch}{1.2} % 可以调节, 1.2指高度是默认的1.2倍
\caption{Ablation study results, in terms of BD-Rate (\%) comparison measured with PSNR. }\label{tab:table3}%The baseline without the proposed modules is FVC\cite{Hu2021FVCAN}.
\centering
\begin{tabular}{ m{1.53cm}<{\centering} m{1.53cm}<{\centering} m{1.53cm}<{\centering}  m{1.53cm}<{\centering} m{1.53cm}<{\centering} m{1.53cm}<{\centering} m{1.53cm}<{\centering} m{1.53cm}<{\centering} }
\toprule
RDT & MGP & SFD-T & HEVC B & HEVC C & HEVC D & HEVC E & Average \\ \hline
\ding{55} & \ding{55} & \ding{55} & 0.0 & 0.0 & 0.0 & 0.0 & 0.0 \\%\hline
\ding{51} & \ding{55} & \ding{55} & -17.8 & -15.3 & -18.2 & -24.2 & -18.9 \\%\hline
\ding{51}& \ding{51} & \ding{55} & -41.4 & -44.0 & -35.3 & -53.3 & -43.5 \\%\hline
\ding{51} & \ding{51} & \ding{51} & -67.0 & -71.8 & -61.1 & -75.1 & -68.7  \\
\bottomrule
\end{tabular}
\end{table*}

\begin{figure*}[t]
	\centering
\begin{minipage}[b]{0.5\linewidth}
		\centering
		\subfloat[A sample frame]{\includegraphics[width=\textwidth]{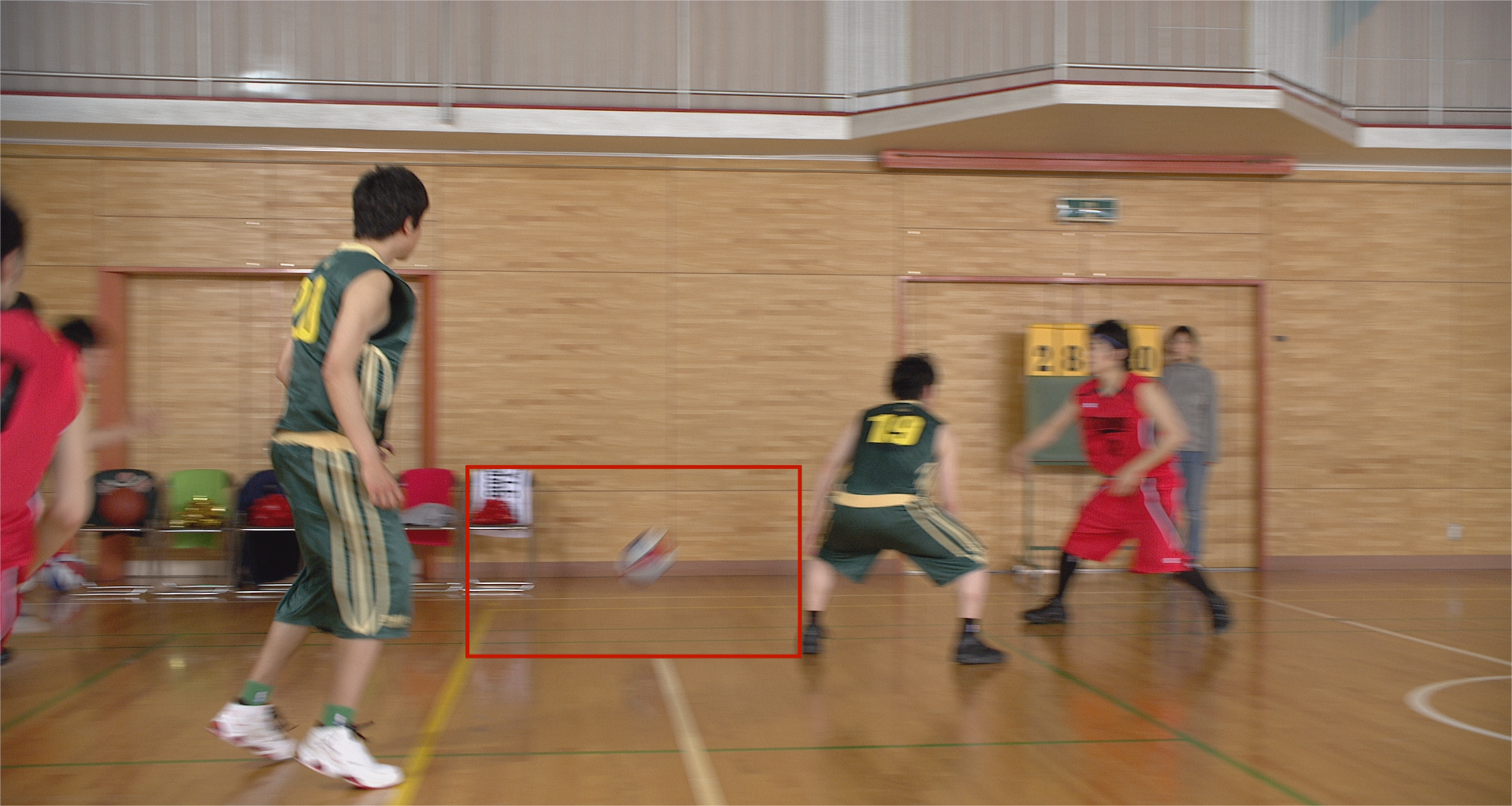}}\label{fig_6a}
	\end{minipage}
	\begin{minipage}[b]{0.44\linewidth}
		\centering
  \subfloat[Ground Truth]{\includegraphics[width=.46\textwidth]{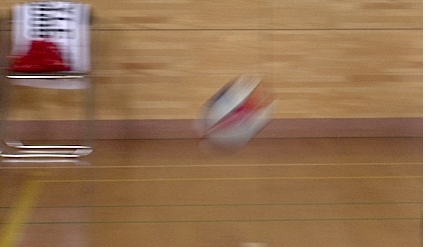}\label{fig_6b}}\hspace{3pt}
  \subfloat[FVC prediction]{\includegraphics[width=.46\textwidth]{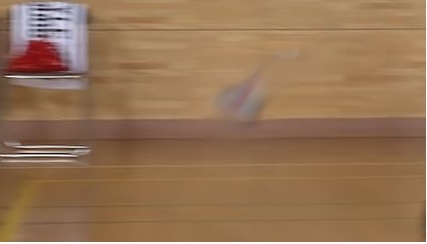}\label{fig_6c}}\\
		\subfloat[Coarse prediction]{\includegraphics[width=.46\textwidth]{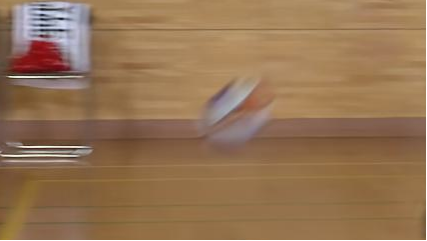}\label{fig_6d}}\hspace{3pt}
		\subfloat[MGP prediction]{\includegraphics[width=.46\textwidth]{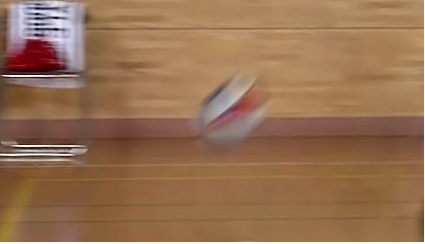}\label{fig_6e}}
	\end{minipage}
	\caption{Visual results comparison.}
 \label{fig_6}
\end{figure*}

\subsection{ Comparison with State-of-the-art Methods}
To evaluate the performance, the proposed method is compared with the existing State-of-the-art methods, including HM-16.25 \cite{HM}, VTM-17.0 \cite{VTM}, ECM-5.0 \cite{ECM}, FVC \cite{Hu2021FVCAN}, DCVC \cite{Li2023NeuralVC}, DCVC-HEM  \cite{Li2022HybridSE}, and DCVC-TCM \cite{Sheng2021TemporalCM}. Among these methods, HM, VTM and ECM are the traditional block-based video compression methods while the others are LVC methods. The results are shown in Table \ref{tab:table1} and Table \ref{tab:table2} for PSNR and MS-SSIM based BD-Rate comparison, respectively. The result of VTM is used as anchor for the BD-Rate calculation. From Table \ref{tab:table1}, it can be seen that the proposed method achieves better performance than the existing LVC methods. It obtains an average bitrate saving of 13.5\% compared to the anchor, while the other LVC methods perform worse (or similar) than VTM. The RD curve comparisons are shown in Fig. \ref{fig_5}, where our method performs best compared to the others.
\vfill

\subsection{ Ablation Study}
Ablation experiments are further conducted to validate the effectiveness of each proposed module in our method. The baseline without the proposed modules is the FVC model \cite{Hu2021FVCAN} and the HEVC dataset is used for evaluation. 

{\bf RDT based motion estimation and compensation.} On top of the baseline FVC, RDT is first used to replace the CNN-based motion estimation and compensation component. The result is shown in Table \ref{tab:table3}. It can be seen that compared with the baseline FVC, significant improvement in terms of BD-Rate, 18.9\% reduction on average, is achieved with our RDT module. This indicates that the proposed RDT based motion estimation method can obtain higher-quality motion information from color space. As shown in  Fig. \ref{fig_6}, a comparison between Fig. \ref{fig_6c} and Fig. \ref{fig_6d} shows that the ball in our coarse-grained prediction picture obtained with RDT are more clear and complete than the FVC prediction. This further demonstrates the advantages of the proposed RDT in capturing motion details and stably transforming features between color space and geometric space.

{\bf MGP based on multi-frame enhancement}. MGP is further added on top of the FVC and RDT to validate the effectiveness of using multi-reference frame to enhance the prediction generation. As shown in Table \ref{tab:table3}, the performance is further improved by 24.6\% in terms of BD-rate. This demonstrates the effectiveness of using MGP to explore the multi-reference information for prediction frame refinement. With sufficient temporal information, the quality of prediction frame is further improved. A visual result comparison in Fig.\ref{fig_6d} and Fig.\ref{fig_6e}  illustrates that the prediction frame with our MGP module contains more temporal motion details and sharper edges compared to the FVC prediction and coarse-grained prediction.

{\bf SFD-T based residual compression.}  The performance of the proposed SFD-T module can be further observed by comparing the results of the RDT \& MGP  and the result of the full module in row 2 and row 3 of Table \ref{tab:table3}, respectively. It can be observed that the proposed SFD-T provides a significant 25.2\% BD-Rate gain over the CNN based residual compression in FVC. This demonstrates that the SFD-T module effectively reduces redundancy in the spatial feature distribution embedded in the temporal prediction, improving the residual compression efficiency. Notably on HEVC test sets, the proposed method with all the modules surpasses the baseline FVC by an average of 68.7\% BD-Rate saving, demonstrating the effectiveness of our method. 

\section{Conclusion}\label{sec5}
In this paper, we propose a novel Spatial-Temporal Transformer-based video compression (STT-VC) framework, containing RDT, MGP and SFD-T, designed for inter-frame coding. RDT is developed for high-quality motion estimation and compensation by producing a stable feature transformation from color space to geometric motion space. MGP further enhances the prediction feature by exploring the multi-reference frame information. It takes full advantage of the coarse-grained prediction generated by the RDT and characterized by the coded motion information.  Then SFD-T is designed to improve residual compression efficiency by leveraging joint temporal-spatial feature distribution priors to further reduce spatial redundancy in the residuals. Experimental results demonstrate that the proposed STT-VC framework outperforms VTM by 13.5\% on average in all tests and achieves the best performance. Ablation studies confirm the effectiveness of each proposed module in the framework and achieve a coding gain of 68.7\% BD-Rate saving against the baseline.

{\small
\bibliographystyle{ieee}
\bibliography{refer}
}

\end{document}